\documentstyle[aps,preprint]{revtex}
\begin{document}
\draft
\title{The Axial Current in Electromagnetic Interaction}
\author{ Myung Ki Cheoun \footnote{e.mail : cheoun@phya.yonsei.ac.kr}, 
K.S.Kim, Il-Tong Cheon }
\address
{Department of Physics, Yonsei University,
Seoul, 120-749, Korea \\
(November, 3, 1998)}
\maketitle
\begin{abstract}
We discussed the possibility that 
the charged axial currents of matter fields 
could be non-conserved 
in electromagnetic interaction at $O(e) $ order. It means that chiral symmetry
is broken explicitly by electromagnetic interaction. This explicit 
symmetry breaking of chiral symmetry is shown to lead the mass differences 
between the charged and neutral particles of matter fields. 
\end{abstract}

\vspace{1cm}
\pacs{PACS numbers : 24.80+y, 13.75.Gx, 13.40.Dk }

Axial current is Noether current of chiral symmetry, which
is a good symmetry in 2 flavour, i.e., SU(2) symmetric world, but
broken more or less in SU(3) symmetric world.
The global nature of the chiral symmetry generates the massless Nambu 
Goldstone (NG) boson through its spontaneous symmetry breaking (SSB) 
\cite{Pe95}. The
pseudo-scalar (PS) bosons are the explicit manifestation of the NG bosons.

In the intermediate energy nuclear physics, however, the PS bosons have finite 
masses by explicit symmetry breaking (ESB) of chiral symmetry 
due to non-vanishing
current quark masses, for the instance of pion, 
$m_{\pi}^2 f_{\pi} = { 1 \over 2} ( m_u
+ m_d ) < {\bar q} q > $, where $< {\bar q} q > = < {\bar u } u > +
<{\bar d} d> $ is the vacuum expectation value (VEV) of quark pairs.
Therefore the conservation of axial current is not maintained any more, 
so that one presumes the partial conservation of the axial current
(PCAC), $\partial_{\mu} A_{\mu}^{a} = - f_{\pi} m_{\pi}^2 {\pi}^a $. 
But the SSB property of the vacuum is still preserved 
even in the ESB because the ESB is included
perturbatively in order to retain the SSB property of vacuum.

On the other hand, electro-magnetic (EM) interaction with matter fields
are usually introduced as an external U(1) local gauge field.
The constraint of gauge invariance gives well defined 
interaction Lagrangian of the matter fields with EM fields.

Naturally one expects that the EM interaction Lagrangian influences 
on the axial current of the system, for instance,
$\pi - N $ system in this letter.
We show that i) the axial current is not conserved any more 
in EM interaction. Therefore, the Lagrangian modified by external
EM interaction causes another ESB mechanism of chiral symmetry
at $ O(e) $ order besides the ESB due to finite pion mass. 
ii) This ESB gives rise to the
mass differences among the matter fields.

Before going to further discussions, we note
two following facts : i) One has the anomalous non-conservation
of the axial current beyond the classical field theory, 
known as the Adler-Bell-Jakiew (ABJ) anomaly \cite{Ad69}. 
But this anomaly appears at $O(e^2)$ order although
it has fruitful physical consequences. ii) One is used to the concept
of `` minimal coupling scheme `` in the EM coupling to the matter 
fields, which is widely used in the nuclear physics relevant to the 
photon reaction. 
But this scheme is not justified in the following sense. 
Given the Lagrangian of matter fields, one derives (axial) current 
and includes the EM interaction by the minimal coupling, i.e., replaces 
$\partial_{\mu}$ in the axial current 
by $\partial_{\mu} - ie B_{\mu}$ where $B_{\mu}$ is photon field and
$e$ is the EM coupling constant. 
But one could not be sure
that the current modified this way equals to the current obtained
directly from the Lagrangian including the EM interaction in gauge
invariant way. 

Since the above arguments are well reflected in the linear $\sigma$ model,
we start from the Lagrangian of the following linear $\sigma$ model 
${\cal L}_{\sigma} = 
{\cal L}_N + {\cal L}_{\pi N} + {\cal L}_{\pi} $ 
with

\begin{eqnarray}
{\cal L}_N & = & {\bar \Psi}  i \gamma^{\mu} \partial_{\mu}  \Psi ~~,~~
{\cal L}_{\pi N} = {\bar \Psi} 
 g ( \sigma + i {\vec \tau} \cdot {\vec \pi} \gamma_5 ) \Psi
~, \\ \nonumber
{\cal L}_{\pi} & = & { 1 \over 2} [ {( \partial_{\mu} {\vec \pi} )}^2 +
{( \partial_{\mu}{\sigma} )}^2 ]
 - { 1 \over 2} {\mu}^2 ( {\vec \pi}^2 + {\sigma}^2 )
- { {\lambda} \over 4} {( {\vec \pi}^2 + {\sigma}^2 )}^2~.
\end{eqnarray}
Although this Lagrangian is invariant under SU(2) chiral transformations, 
$ \Psi \rightarrow {\Psi}^{'} =  exp (  i {\gamma}_5 {\vec \eta} \cdot
{{\vec \tau} \over 2} ) \Psi , 
{\bar \Psi} \rightarrow {{\bar \Psi}}^{'} 
= {\bar \Psi} exp (  i {\gamma}_5 {\vec \eta} \cdot
{{\vec \tau} \over 2} ) $, ${\vec \pi} \rightarrow {\vec \pi}^{'} = 
{\vec \pi} - \sigma {\vec \eta}$, and, $ \sigma \rightarrow \sigma^{'} =
 \sigma + {\vec \pi} \cdot {\vec \eta} $, it describes only 
massless fermion and massless bosons. Later on we will show 
the mass generation of the matter fields 
and its modification by the EM interaction. 

To generate the EM interactions ( external U(1) gauge
field ) with the fermion fields
we firstly consider ${\cal L}_N$ and introduce 
photon field $B_{\mu}$ via the covariant derivative $D_{\mu} \Psi
= (\partial_{\mu} + i V_{\mu} ) \Psi $, where 
$V_{\mu} = - e T_3 B_{\mu}$ with $T_i = {\tau_i / 2}$. It has the following
transformation rule in order to be invariant under
a local transformation $h(x)$, i.e., $D_{\mu}^{'} \Psi^{'} =
h D_{\mu} \Psi$,
\begin{equation}
\delta V_{\mu} = i ( \partial_{\mu} h ) h^{-1} - [ V_{\mu}, h ] h^{-1} 
~~.
\end{equation}
Since the Gell-Mann Levy method is a very convenient tool 
to derive the current
from the given Lagrangian, we assume temporally the local chiral 
transformation ($\eta = \eta (x)$) to extract axial current \cite{Bh88}. 
The change of ${\cal L}_N$ in the local SU(2) 
chiral transformation, 
$ \delta {\cal L}_N = {\bar \Psi}^{'}  i \gamma^{\mu} D_{\mu}  \Psi^{'} -
 {\bar \Psi}  i \gamma^{\mu} D_{\mu}  \Psi $,  is given by
\begin{equation}
\delta {\cal L}_N =
i {\bar \Psi} [ i \gamma^{\mu} ( \partial_{\mu} {\vec T} \cdot {\vec \eta} )
\gamma_5 - 
i \gamma_{\mu} ( e B^{\mu} \epsilon_{ 3 l c} T_l \eta_c ) \gamma_5
 ] \Psi ~.
\end{equation}
Using the Gell-Mann Levy equations 
we get the following axial current and its divergence 
\begin{eqnarray}
A_{c}^{\mu} (N) = - {  { \partial \delta {\cal L}_N } \over { \partial ( \partial_{\mu}
\eta_c)}} = {\bar \Psi } \gamma^{\mu} T_c \gamma_5 \Psi \\ \nonumber
\partial_{\mu} A_{c}^{\mu}  (N) = { {\partial \delta {\cal L}_N } \over
{\partial \eta_c}} = e \epsilon_{3 l c} {\bar \Psi} {B \hspace{-0.08in}/} 
T_l \gamma_5 \Psi  ~.
\end{eqnarray}
The EM interaction does not affect the axial current itself, but breaks 
its conservation apart from the case of neutral axial current, where
the ABJ anomaly plays a role in breaking 
the conservation of the axial current at $O(e^2)$ order.

The transformation rule of the mesons comes from retaining the 
interaction Lagrangian ${\cal L}_{\pi  N}$ chiral invariant, so that
${\vec \phi} \rightarrow
{\vec \phi}^{'} = {\vec \phi} - f {\vec \eta}$. ${\vec \phi}$
is related to $\sigma $ and ${\vec \pi}$ fields through chiral field
$U = exp ( i {\vec \phi} \cdot {\vec T} / f ) = { 1 \over { 2 f}}
( \sigma + i {\vec \tau} \cdot {\vec \pi} )$ which transforms
as $U \rightarrow U^{'} = U g $ with $g = exp ( - i {\vec T } \cdot 
{\vec \eta (x)})$.

For the meson sector,
\begin{equation}
{\cal L}_{\pi} = { 1 \over 2} [ {( \partial_{\mu} \sigma)}^2
+ {(\partial_{\mu} {\vec \pi})}^2 ] = f^2 Tr ( \partial_{\mu} U 
\partial^{\mu} U^{+} )~,
\end{equation}
we make the similar procedures to
take the EM interaction with $\phi$ fields into account. Exploiting
the covariant derivative for $U$ field, 
$D_{\mu}^{\pi}  U = \partial_{\mu} U + i [ U, V_{\mu} ] $,
we find the following transformation rule of $V_{\mu}$ to be invariant
under a local transformation $h$, i.e. $D_{\mu}^{'} U^{'} {D^{\mu}}^{'} U^{'}
= D_{\mu} U D^{\mu} U$,
\begin{equation}
\delta V_{\mu} = - i ( \partial_{\mu} h^+ ) h - [ V_{\mu}, h^+ ] h ~.
\end{equation}
The change of ${\cal L}_{\pi}$ under transformation $g$, 
$\delta {\cal L}_{\pi}
= f^2 Tr( D_{\mu} U^{'} D^{\mu} U^{'} ) - f^2 Tr ( D_{\mu} U D^{\mu} U )$, 
is represented as 
\begin{eqnarray} 
\delta {\cal L}_{\pi}&  = & f^2 Tr (
\alpha_{\mu} {\beta^{\mu}}^{+} + \beta_{\mu} {\alpha^{\mu}}^{+}
 + i \beta_{\mu} V^{\mu} + i V_{\mu} {\beta^{\mu}}^{+} \\ \nonumber
& & + i ( \alpha_{\mu} - V_{\mu}) 
[  { { {\vec T} \cdot {\vec \phi} } \over f} , {\beta^{\mu}}^{+} ]
 + i V_{\mu} O^{\mu} ( \phi^2 ) - \alpha_{\mu} [ V_{\mu} , {\vec T} \cdot
 {\vec \eta}  ] ~)~,
\end{eqnarray}
where $ \alpha_{\mu} = { 1 \over f} \partial_{\mu} 
( {\vec T} \cdot {\vec \phi} ) $, $\beta_{\mu} = - i \partial_{\mu}
(  {\vec T} \cdot {\vec \eta} ) $ and local chiral transformation 
( $\eta = \eta (x)$ ) is also assumed.
Upto $O ( V_{\mu}, {\phi}, \eta , e )  = O (1) $ order, we obtain
\begin{eqnarray}
A_{c}^{\mu}  ( \pi) & = &  -  {  {\partial \delta {\cal L }_{\pi}  } \over { \partial  ( \partial_{\mu }
\eta_c ) } } = f \partial^{\mu} \phi_c + f e B^{\mu} \epsilon_{3 l c}
\phi_l + O^{\mu} ( 2 ) \\ \nonumber
\partial_{\mu} A_c^{\mu} ( \pi) & = & f e \epsilon_{3 l c} B^{\mu} 
\partial_{\mu} \phi_l  + O (2 ) ~. 
\end{eqnarray}
Likewise to the nucleon case, the conservation of the axial current is 
broken at $O(e)$ order. 
But in the axial current the EM effects showed up directly.

Finally, total axial current and its divergence are 
changed by the EM interaction in the following way
\begin{eqnarray}
A_c^{\mu} & = & {\bar \Psi } \gamma^{\mu} T_c \gamma_5 \Psi +
f \partial^{\mu} \phi_c + f e B^{\mu} \epsilon_{3 l c} \phi_l + O (2 )
\\ \nonumber
\partial_{\mu} A^{\mu}_c & = & e \epsilon_{ 3 l c} ( {\bar \Psi} {B \hspace{-0.09in}/} 
T_l \gamma_5 \Psi  + f B^{\mu} \partial_{\mu} \phi_l ) + O ( 2 )  \\ \nonumber
& = & e \epsilon_{3 l c} B_{\mu} A^{\mu}_l + O (2 ) ~.
\end{eqnarray}
This form can be reduced to the normal form if $e$ goes to zero.

The above form can be expressed into the more familiar representation
known as "the minimal coupling scheme" in current level, if we introduce
the generalized covariant derivative  $D_{\mu ( \pm) } = \partial_{\mu}
 \mp i e B_{\mu}$ in the total axial current
$A^{ (\pm) \mu} = A^{ (\pm) \mu} (N) + A^{ (\pm) \mu} (\pi)$,
\begin{eqnarray}
D_{\mu( \pm) } A^{ (\pm) \mu} & = &  0 ~,\\ \nonumber
A^{ (\pm) \mu} (N) & = & {\bar \Psi} \gamma^{\mu} \tau_{\pm} \gamma_5 \Psi~,~~
A^{ (\pm) \mu} ( \pi) = f ( \partial^{\mu} \mp i e B^{\mu} ) \phi_{\pm}~,
\end{eqnarray}
where $\tau_{\pm} = { 1 \over 2} ( \tau_1 \pm i \tau_2 )$ and $ \phi_{\pm} =
{ 1 \over {\sqrt 2}} ( \phi_1 \pm i \phi_2 ) $. This result can be also derived by exploiting the
generalized Euler equation \cite{Ad65} 
and applied to the quenching problem of induced pseudo-scalar
coupling constant in the radiative muon capture \cite{Ch98}.

Non-conservation of the axial current as shown above
means just that the chiral symmetry is broken explicitly by EM interaction
at $O(e) $ order. 
Now let's discuss the physical meaning of the ESB by the EM interaction. 
Actually the chiral symmetry is a global, 
so that we have to switch off the locality of chiral symmetry. Therefore
the change of ${\cal L}_{\sigma}$ due to EM interaction is finally 
given at $O(e)$ order,
\begin{equation}
\delta {\cal L}_{\sigma} = {\cal L}_{\sigma}^{e.m.} - {\cal L}_{\sigma}
= {\bar \Psi} \gamma_{\mu} B^{\mu} \gamma_5 T_l \Psi +
 e B^{\mu} \epsilon_{3 l c} \phi_l \partial_{\mu} \phi_c + O(e^2) ~.
\end{equation}
Usually, 
in order to create the pion mass, 
the chiral ESB term $\zeta\sigma $ 
should be included into ${\cal L}_{\sigma}$. By exploiting the Gell-Mann Levy
method, similarly to the above, one sees $\partial_{\mu} A_c^{\mu} = - \zeta 
\pi_c$. 
The $\zeta \sigma$ term is to generate the $\pi^0$ mass,
$m_{\pi^0}^2 = \zeta / < \sigma >$. Here $< \sigma > $ is the mean field 
taken in the degenerate vacuum, as a result the vacuum is spontaneously
broken, to introduce new field $\sigma^{'} = \sigma + < \sigma >$ which is
vanishing at the ground state.
On the other hand, the fermion matter field acquires a mass $M_N = - g 
< \sigma  > $. Therefore $\zeta$ is finally given as $ - m_{\pi^0}^2 M_N / g$.

Now let us switch on the EM interaction. Then $\zeta$ is changed as $\zeta^{'}$
which includes effectively the additional ESB term ( see eqs. (8-9) and 
(11) ). Here we would lik to remind that $\zeta$ is introduced perturbatively 
in order not to change the SSB property of vacuum. In this context, 
$\zeta^{'} = \zeta + \delta \zeta ( e )$ can be also safely introduced 
perturbatively because $\delta \zeta (e ) $ is much smaller than $\zeta$. 
Similarly, $ < \sigma > \rightarrow < \sigma^{'} >$.
Finally, since we can conjecture that the masses of charged particles are
obtained from those of neutral particles with additional inclusion of EM
interaction we get 
\begin{equation}
\zeta = - m_{\pi^0}^2 { M_N \over g}~~,~~ \zeta^{'} = - m_{\pi^{\pm}}^2
 { M_P \over g }~,
\end{equation}
where $M_N$($M_P$) and $m_{\pi^0}$($m_{\pi^{\pm}})$ mean 
the masses of neutron(proton) and neutral(charged) pions, respectively.
If we assume $\zeta \simeq \zeta^{'}$, we obtain a relation of mass ratio 
\begin{equation}
{( { m_{\pi^0}  \over m_{{\pi}^{\pm}} }  )}^2 = { M_P \over M_N}~.
\end{equation}
This relation shows that the charged particle mass is heavier 
than neutral particle in case of meson, while in case of
fermion neutral particle is heavier than charged particle. But,
actually this relation is deviated about 7 \% if we use the measured values. 
The mass difference
in pion can be explained by the virtual photon propagation, which comes 
from the interaction of 
the 2nd term in eq.(11), but the mass difference in nucleon cannot be 
fully explained only by EM interaction. 
Therefore the 7 \% deviation should be attributed to the
effects at quark level of hadrons.

Finally we make a brief summary. If we introduce the EM interaction
to the matter fields in U(1) gauge invariant way, the chiral symmetry 
is broken explicitly in the total Lagrangian, so that the axial current is
not conserved any more. This ESB of chiral symmetry just gives
the mass differences between the charged and neutral particles 
in the matter fields. In specific, such mass difference in the pion sector is
just reversed in case of the nucleon, namely, $m_{\pi^{\pm}} > m_{\pi^{0}}$
while $M_N > M_P$. We showed a mass relation among the particles. 
The small deviation in the nucleon side should
be understood at quark level.

\vspace{1.5cm}

{\bf Acknowledgement}

The work of Il-Tong Cheon was by the
Basic Science Research Institute Program, Ministry of Education of
Korea, No. BSRI-97-2425.

\end{document}